\documentclass[12pt,showkeys]{revtex4-1}

\usepackage{bm}

\usepackage{graphicx}
\usepackage{subfig}

\expandafter\let\csname equation*\endcsname\relax
\expandafter\let\csname endequation*\endcsname\relax
\usepackage{amsmath}
\usepackage{color}
\usepackage{amssymb,amsfonts}
\usepackage{color,psfrag}
\usepackage{mathrsfs}

\newcommand{\nn}{\nonumber}
\newcommand{\bb}{\begin{equation}}
\newcommand{\ee}{\end{equation}}

\newcommand{\eref}[1]{equation~(\ref{#1})}
\newcommand{\efig}[1]{figure~\ref{#1}}
\renewcommand{\Re}{{\rm Re}}

\newcommand{\ftwo}{\makebox{\small $\frac{1}{2}$}}
\newcommand{\fitwo}{\makebox{\small $\frac{i}{2}$}}

\renewcommand{\d}{{\rm d}}

\newcommand{\D}{{\mathscr D}}
\newcommand{\bD}{{\mathscr D}^{\dagger}}
\newcommand{\vk}{\Psi}

\newcommand{\ff}[2]{\frac{#1}{#2}}
\newcommand{\e}{e}
\newcommand{\dQ}{Q^{\dag}}

\newcommand{\va}{\bm{\alpha}}

\newcommand{\vs}{\bm{\sigma}}
\renewcommand{\Im}{\,{\rm Im}}

\newcommand{\trans}{\top}

\newcommand{\he}{h}
\newcommand{\diag}{\textrm{diag}}
\newcommand{\nm}{N}
\newcommand{\Ai}{{\mathscr A}}

\newcommand{\phase}[1]{\gamma_{#1}}
\newcommand{\Fc}[1]{F(#1)}
\newcommand{\DN}{\Delta \rho}
\newcommand{\zd}{w}
\newcommand{\GE}{G}

\newcommand{\xn}{\rho}
\newcommand{\drho}{\Omega}

\begin{document}

\title{Charge oscillations in a simple model of interacting magnetic orbits}
\author{Jean-Yves Fortin}

\affiliation{Department of Physics and Astronomy and Center for Theoretical Physics,
Seoul National University, Seoul 08826, Korea 
\footnote{Permanent address: Laboratoire de Physique et Chimie Th\'eoriques, CNRS UMR 7019, Universit\'e de Lorraine, F-54000 Nancy, France}}

\email{jean-yves.fortin@univ-lorraine.fr}

\begin{abstract}
Exact eigenstates for a set of two or more interacting electronic orbits in a magnetic field are studied for 
a class of factorized Hamiltonians with coupled Fermi surfaces. We study the condition for the existence
of annihilation-creation operators that allows for the construction of eigenstates. For the case of two interacting
cyclotronic orbits, we consider the oscillations of the overlap function and the transfer of charge density
between the orbits as function of the inverse field. The expressions of the Fourier frequencies are given in the semiclassical regime and they depend on the geometrical structure of the electronic bands. A generalization of this construction is provided for a chain of several interacting orbits with exact eigenfunctions. 
\end{abstract}


\maketitle
\section{Introduction}
\begin{figure}%
\centering
\includegraphics[width=0.8\columnwidth,clip,angle=0]{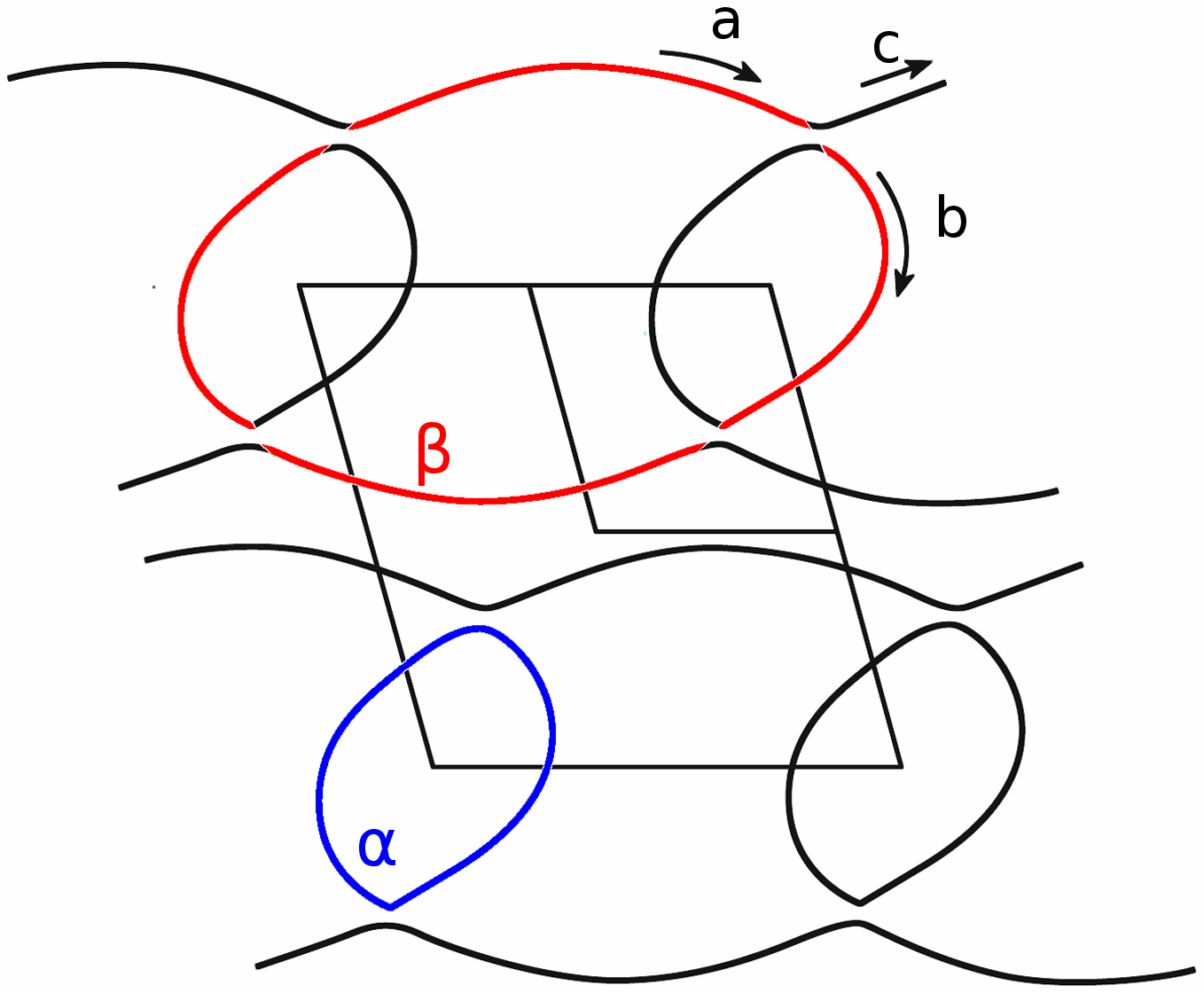}%
\caption{Fermi surface of the organic conductor (BEDO-TTF)$_5$[CsHg(SCN)$_4$]$_2$, from \cite{Lyubovskii:2004,Fortin:2017}. 
It is composed of a chain of small orbits $\alpha$ and magnetic breakdown orbits $\beta$. Electron trajectories 
are represented by incoming waves (a) that are scattered in (c) and transmitted in (b).
}%
\label{fig0}%
\end{figure}

The two-state problem in quantum mechanics, such as the Rosen-Zener two-level model, gives precise information on how the wavefunction propagates through a junction and and has important applications in time-driven quantum systems \cite{Bambini:1981,inbook:Hanggi1998,Torosov:2008,Torosov:2011}, where resonance and phase shifts are studied in details for exact solvable cases of potentials. This general problem can be applied in the case of magnetic breakdown as well, where a quasi-particle orbits a Fermi surface which can be composed of multiple sheets connected by junctions through which the particle can tunnel.
A realization of such Fermi surfaces, resembling a linear chain of coupled orbits, is presented in
\efig{fig0} for the organic conductor (BEDO-TTF)$_5$[CsHg(SCN)$_4$]$_2$ \cite{Lyubovskii:2004} 
(BEDO-TTF is the abbreviation for bis-ethylenedioxi-tetrathiafulvalene molecule): an incoming wave-packet (a) 
on a giant orbit $\beta$ is transmitted to the small cyclotronic orbit $\alpha$ (b) and reflected onto the same orbit $\beta$ (c) within the chain. Fermi surfaces with a finite number of interacting orbits also exist and the Fourier spectrum of their magnetic oscillations have been studied in details. For example, compensated Fermi structures with only three bands, made of one hole and two electron pockets, can be found in compound $\alpha$-'pseudo-$\kappa$'-(ET)$_4$H$_3$O[Fe(C$_2$ O$_4$)$_3$]·(C$_6$H$_4$Br$_2$), where the $\alpha$-type and 'pseudo-$\kappa$'-type are conducting and insulating layers respectively \cite{Audouard:2014}.
At the magnetic breakdown junction, the Hamiltonian can be linearized and the two sheets hybridized with some energy coupling $g$. This simplest form of two-level Hamiltonian was solved by Rosen and Zener in a different context \cite{Rosen:1932} using this approximation around the tunneling region. The probability of tunnelling is exponentially small in the ratio between a breakdown field and the magnetic field
\cite{Chambers:1968}, and includes a field dependence of the Onsager phase \cite{Slutskin:1967,Kochkin:1968}, in addition to the $\pi/2$ phase each time the particle is reflected at the junction. 

In this paper we would like to construct simple and exact eigenstates for multi-band conductors that incorporate magnetic tunneling on such Fermi surfaces made for example of a chain of two or more coupled orbits \cite{Pippard:1962}.

In general there exist several possible techniques for solving Hamiltonians with complex potentials, in addition to the semi-classical techniques provided by WKB method \cite{Kaganov:1983,Kadigrobov:1992} dealing with scattering matrices for interacting orbits.  
For example, the factorization approach, or Darboux method, \cite{Infeld:1951,Gendenshtein:1985,Casahorran:1995,Cooper:1995,book:Cooper,Fellows:2009} consists in considering an Hamiltonian which can be factorized as $H_1=\bD\D=-\partial_x^2+V_1(x)$, where $\D=\partial_x+Q(x)$ and $\bD=-\partial_x+Q(x)$ are first order operators. Consider now a partner Hamiltonian $H_2$ constructed by inverting the two operators $\D$ and $\bD$: $H_2=\D\bD=-\partial_x^2+V_2(x)$. $V_1(x)$ and $V_2(x)$ are supersymmetric partner potentials expressed with
$Q(x)$ and its derivative. It follows that the eigenfunctions and eigenvalues of $H_2$ are closely related to those of $H_1$. If $H_1$ is exactly solvable, then $H_2$ is also exactly solvable \cite{Gendenshtein:1985,Cannata:1998,Samsonov:2010}. This allows to construct exact solutions of the Schr\"odinger equation with non-trivial potentials $V_2(x)$, for example for classes of potentials with two wells \cite{Downing:2013}.

Another possibility that we explore in this paper is to consider the same form of factorized Hamiltonian $H=\bD\D$, where $Q(x)$ is now a generalized complex matrix but whose structure is constrained by the condition that the commutator between the operators $\D$ and $\bD$ satisfies $[\D,\bD]=1$. In this case $\D$ and $\bD$ are called ladder or annihilation/creation operators and it is straightforward to construct the eigenstates from the knowledge of the ground state, and the energy spectrum is simply discrete. In the case of a single band, the constraint leads simply to $Q(x)$ being a linear function of $x$ (harmonic oscillator), but for two-band or multi-band systems $Q(x)$ can have a more complicated structure. In addition, we will see that the matrix $Q(x)$ can be chosen such that the two sheets of the Fermi surface of the Hamiltonian present a magnetic gap, which is the purpose of this paper. We will also show that a simple extension to multi-band Hamiltonians leads to a modelization of the linear chain of coupled orbits discussed above and in \efig{fig0}.

%
%
\section{Model}
%
The problem of treating Bloch electrons in a magnetic field has been generally approached using
the Peierls substitution \cite{Peierls:1933,Blount:1962}. The Bloch Hamiltonian for the energy dispersion
$H({\bf k})$ is replaced by $H(-i\nabla+e{\bf A}/\hbar)$, where $-e$ is the charge of the electron. This approximation
is usually valid near the Fermi surface, or for tight-binding Hamiltonians \cite{Alexandrov:1991}.
The quantification of the quasi-momentum $-i\nabla+e{\bf A}/\hbar$ in the Landau gauge ${\bf A}=(0,Bx_1,0)$ leads to an analogy with a one-dimensional particle motion in a potential \cite{Landau9:book} if we identify the dimensionless operators $\hat x=a_2(-i\partial_2+x_1/l_B^2)$ and $\hat p=-ia_1\partial_1$ with effective position and momentum, where $(a_1,a_2)$ are the typical sizes of the Brillouin zone unit cell and $l_B^{-2}=eB/\hbar$ the inverse squared of the magnetic length. These operators
satisfy the commutation relation $[\hat x,\hat p]=ia_1a_2eB/\hbar\equiv i\he$, where $\he=a_1a_2/l_B^2$ is a dimensionless effective Planck constant. It can also be identified as the ratio between the magnetic flux though the unit cell and the quantum flux $h/e$. Usually it is considered as a small control parameter in the semiclassical theories if the cell units in the Brillouin zone are of the order of the Angstr\"om, or if the magnetic field is of the order of the Tesla. We should notice 
that in the case of positive charges, $e$ is instead negative and to keep
the effective Planck constant $\he$ positive we can invert the role of $\hat x$ and $\hat p$ to keep the same
commutation relation.
Starting with the Hamiltonian expression for an electron in a uniform magnetic field $B$, where $H({\bf k})$ is 
quadratic in momenta ${\bf k}$ and describing a simple harmonic oscillator near the Fermi surface which can be therefore factorized in terms of ladder or creation and annihilation operators \cite{book:Yoshioka}, we introduce the representation $\hat p=-i\he\partial_x$ and consider a class of differential operators for multiband systems defined by
\bb\label{D_def}
\bD=\ff{1}{\sqrt{2\he}}\left [-\he\sigma_0\partial_x+\dQ(x)\right ],
\;\D=\ff{1}{\sqrt{2\he}}\left [\he\sigma_0\partial_x+Q(x)\right ]
\ee
from which is constructed the factorized Hermitian Hamiltonian $\hat H$ 
\bb\label{Ham}
\hat 
H=2\he\bD\D=-\sigma_0\he^2\partial^2_x+[\dQ(x)-Q(x)]\he\partial_x-\he Q'(x)+\dQ(x) 
Q(x)
\ee
$Q(x)$ is a general matrix depending on $x$, and $\sigma_{i=0,\cdots,3}$ are 
the Pauli matrices. From a supersymmetric point of view \cite{Gendenshtein:1985}, it is known that the product $\dQ(x)Q(x)$ describes interaction between bosons, whereas the term $\he Q'(x)$ deals with fermion-boson interaction,
both being dependent of the same function $Q(x)$. The fermion-boson interaction is absent if $Q'(x)$ is constant, as for the bosonic harmonic oscillator.
The classical version $H_c$ of the Hamiltonian in the limit $\he\rightarrow 0$ is 
obtained by replacing the operator $\hat p$ by momentum variable $p$
\bb
H_c(x,p)=(\dQ(x)-i\sigma_0p)(Q(x)+i\sigma_0p)=\sigma_0p^2+\dQ(x)Q(x)+ip[\dQ(x)-Q(x)]
\ee
The Fermi surface $H_c(x,p)=E\sigma_0$ in the plane $(x,p)$ is given by the eigenvalues of $H_c$ at constant energy $E$, and is generally composed of two closed surfaces. In the following we
consider examples of matrices $Q(x)$ that lead to magnetic tunneling between the two sheets of the Fermi surface. The commutation relation between operators $\D$ and $\bD$ is given by matrix $P(x)$
\bb\label{commutator}
[\D,\bD]=\ff{1}{2}\left (\partial_xQ(x)+\partial_x\dQ(x)\right 
)+\ff{1}{2\he}[Q(x),\dQ(x)]=P(x)
\ee
If $P(x)$ is the unity matrix, this relation is the usual 
bosonic commutator for creation and annihilation operators $\bD$ and $\D$ 
respectively. And the eigenstates $\vk_n=(\phi_n,\tilde\phi_n)^\trans$ are constructed from the ground 
state $\vk_0$ such that $\D\vk_0=0$, by successively acting the ladder operator 
$\bD$ on $\vk_0$. This is, for example, the case when the system is composed of 
two independent harmonic oscillators, centered at positions $\pm 
x_c$
\bb\label{harmonic}
Q(x)=\dQ(x)=
\left (
\begin{array}{cc}
x+x_c & 0 \\
0 & x-x_c
\end{array}
\right ),\; P=\left (
\begin{array}{cc}
1 & 0 \\
0 & 1
\end{array}
\right )
\ee
We would like the study the condition for which tunneling can occur when $P(x)=1$ for general matrices
$Q(x)$. We first decompose $Q(x)$ in the base of the Pauli matrices 
$Q(x)=\sum_{i=0}^3\alpha_i(x)\sigma_i$, with $\alpha_i$ complex functions, and for which the commutator $P(x)$ is equal to
\bb
P(x)=\ftwo\partial_x(\alpha_0(x)+\bar\alpha_0(x))\sigma_0
+\ftwo\partial_x(\va+\bar\va).\vs+\frac{i}{\he}(\va\times\bar\va).\vs
\ee
Functions $\alpha_i$ are then chosen such as $P(x)=\sigma_0$. This leads to 
a set of differential equations
\bb\label{eq_diff}
\partial_x\Re(\alpha_0)=1,\;\partial_x\Re(\va)+\frac{2}{\he}\Re(\va)\times\Im(\va)=0,\;
\partial_x\Re(\va)^2=0
\ee
This is similar to a rotational equation of motion of a particle located at $\Re(\va)$ around the angular vector $\bm{\Omega}=(2/\he)\Im(\va)$ on the sphere of constant radius $\parallel\Re(\va)\parallel$, where the dynamical time variable is played by $x$. The solution for $\alpha_0$ is given by $\alpha_0(x)=x+iw(x)$ up to a constant, and where $w(x)$ is any real function of $x$. We will restrict in the following this solution to $\alpha_0(x)=x$, which corresponds to the displacement of the oscillator in the harmonic case. We will see further below that the presence of $w(x)$ is equivalent to add a gauge term $\e^{i\he^{-1}\int^x w(y) dy}$ to the wavefunctions. For the other components,
let us consider the solution $\va=(u(x),v(x),i\omega(x)+x_c)^\trans$ where $u$, $v$, and $\omega$ are functions with real values. The constant $x_c$ gives the location of the center of the two oscillators at $x=\pm x_c$. The system of equations satisfying (\ref{eq_diff}) is given by
\bb
u'(x)+\frac{2}{\he}\omega(x)v(x)=0,\;
v'(x)-\frac{2}{\he}\omega(x)u(x)=0
\ee
whose solutions are chosen such that $u(x)=g\cos \theta(x)$ and $v(x)=g\sin\theta(x)$, with $\theta(x)=2\he^{-1}\int^x\omega(y)dy$ and $g$ a constant (coupling) parameter. Therefore 
\bb\label{sol}
\va=(g\cos\theta(x),g\sin\theta(x),i\omega(x)+x_c)^\trans
\ee
The matrix $Q(x)$ is then expressed as
\bb\label{Qgen}
Q(x)=
\left (
\begin{array}{cc}
x+x_c+i\omega(x) & g\e^{-i\theta(x)} \\
g\e^{i\theta(x)} & x-x_c-i\omega(x)
\end{array}
\right )
\ee
and the classical Hamiltonian is given by
\bb\label{Hex1}
H_c(x,p)=
\left (
\begin{array}{cc}
(p+\omega)^2+(x+x_c)^2+g^2 & 2g\e^{-i\theta}(x-i\omega) \\
2g\e^{i\theta}(x+i\omega) & (p-\omega)^2+(x-x_c)^2+g^2
\end{array}
\right )
\ee
The Fermi surface is given by two sheets, solutions of the equation $\det(H_c-\sigma_0E)=0$
\bb\label{FS}
E=p^2+\omega(x)^2+x^2+x_c^2+g^2\pm 2\sqrt{(p\omega(x)+xx_c)^2+g^2(x^2+\omega(x)^2)}
\ee
%
\begin{figure}%
\centering
\includegraphics[width=0.5\columnwidth,clip,angle=0]{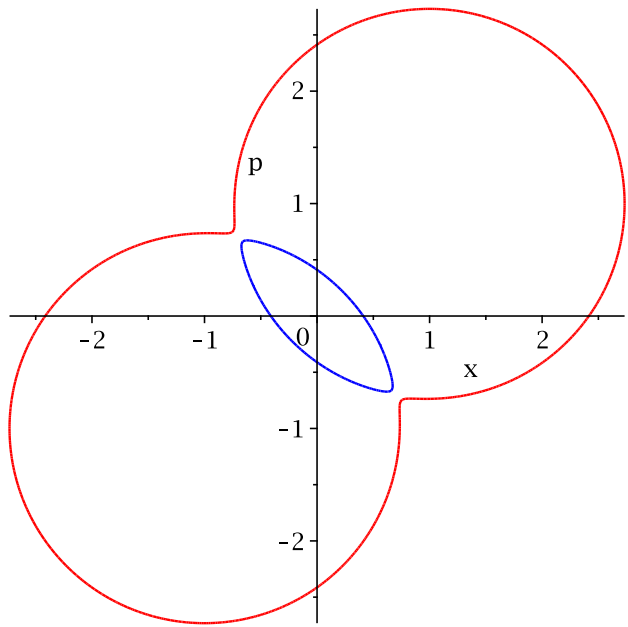}
\caption{Fermi surface of \eref{FS} for parameters $E=3$, $\omega(x)=1$, $x_c=1$, and $g=0.05$. The red surface is given in \eref{FS} by the minus sign, and the blue small pocket by the plus sign.}%
\label{fig1}%
\end{figure}
%
\begin{figure}%
\centering
%
\includegraphics[width=0.5\columnwidth,clip,angle=0]{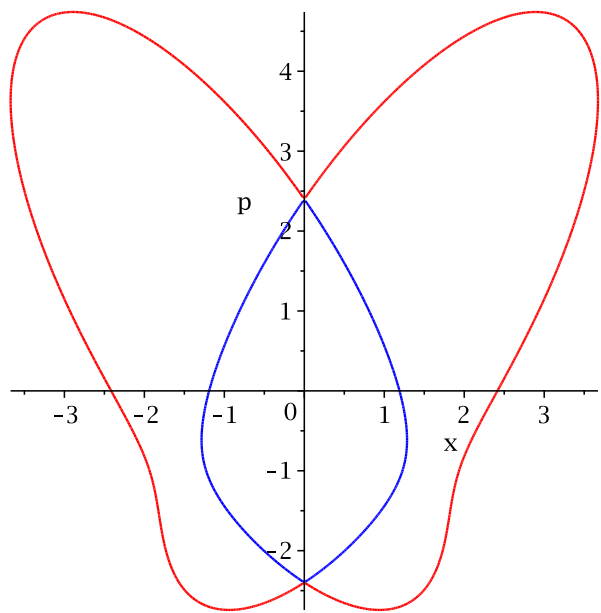}
\caption{Fermi surface of \eref{FS} for parameters $E=5$, $\omega(x)=x$, $x_c=1$, and $g=0.5$. There is no gap in this case as the two sheets intersect at $x=0$.}
\label{fig2}%
\end{figure}
%
As example, we have plotted in \efig{fig1} the surface composed of two sheets at energy $E=3$ with parameters $x_c=1$, $g=0.05$ and function $\omega(x)=1$. The two bands are separated by a gap controlled by $g$. In \efig{fig2} is another example of section with $\omega(x)=x$, $x_c=1$, and $g=0.5$, at energy $E=5$. In the latter case, the two surfaces intersect at $x=0$ because $\omega(0)=0$ and the square root in \eref{FS} is zero. In general if $\omega(0)\ne 0$ the two surfaces do not intersect and a gap appears.

We should notice that $H_c$ depends on $\he$
through the non-diagonal elements $e^{\pm i\theta(x)}$ which become highly oscillating in the classical limit when $\he\rightarrow 0$, but the equation for the Fermi surface (\ref{FS}) in this limit does not depend on $\he$ since the two exponentials in \eref{Hex1} cancel each other. The form of the matrix $Q$ in \eref{Qgen} ensures that the commutator
$P(x)=\sigma_0$ and that the eigenfunctions $\vk_n=(\phi_n,\tilde\phi_n)^\trans$ can be constructed exactly by applying $\bD$ successively on the ground state $\vk_0$: $\vk_n={\bD}^n\vk_0/\sqrt{n!}$, with Landau level energies $E_n=2\he n$. The ground state energy is $E_0=0$, contrary to the usual harmonic oscillator for which the spectrum is $E_n=(2n+1)\he$ and the zero point energy $E_0=\he$. The zero point energy depends on the term $hQ'(x)$ in \eref{Ham} which generates the interaction between bosons and fermions in the supersymmetry theories \cite{Gendenshtein:1985}.

To solve for the eigenfunctions, we introduce the transformation
\begin{align}\nn
Q(x)&=U^{^{\dagger}}(x)Q_0(x)U(x)+hU^{^{\dagger}}(x)\partial_xU(x),
\\ \label{eq_unitary}
Q_0(x)&=
\left (
\begin{array}{cc}
x+\delta & 0\\ 
0 & x-\delta
\end{array}
\right ),\;
U(x)=e^{\fitwo\varphi\sigma_2}e^{\fitwo\theta(x)\sigma_3}
\end{align}
where $U(x)$ is a unitary matrix, and $(\varphi,\delta\ge 0)$ two parameters satisfying the following relations:
\bb
g=\delta\sin(\varphi),\;x_c=\delta\cos(\varphi)
\ee
The Hamiltonian can then be expressed as $H=U^{\dagger}H_0U$, with $H_0=2\he\bD_0\D_0$, $\D_0=(\he\sigma_0\partial_x+Q_0(x))/
\sqrt{2\he}$. The eigenfunctions are therefore related to the elementary pair of harmonic oscillators
with Hermite polynomials
\begin{align}
\vk_n=
\left (
\begin{array}{c}
\phi_n \\ 
\tilde\phi_n
\end{array}
\right )
=\frac{1}{\sqrt{n!2^n}(\pi \he)^{1/4}}U^{\dagger}\left (
\begin{array}{c}
A H_n\left (\frac{x+\delta}{\sqrt{\he}}\right )e^{-\ftwo (x+\delta)^2/h}
\\
B H_n\left (\frac{x-\delta}{\sqrt{\he}}\right )e^{-\ftwo (x-\delta)^2/h}
\end{array}
\right )
\end{align}
where $A$ and $B$ are complex constants. $U^{\dagger}$ takes explicitly the form
\begin{align}
U^{\dagger}=\left (
\begin{array}{cc}
\cos\left (\frac{\varphi}{2}\right )e^{-\fitwo\theta(x)}&
-\sin\left (\frac{\varphi}{2}\right )e^{-\fitwo\theta(x)}
\\ 
\sin\left (\frac{\varphi}{2}\right )e^{\fitwo\theta(x)}&
\cos\left (\frac{\varphi}{2}\right )e^{\fitwo\theta(x)}
\end{array}
\right )
\end{align}
and the values of the components of $\vk_n$ are simply given by
\begin{align}\nn
\phi_n(x)&=\frac{e^{-\fitwo\theta(x)}}{\sqrt{n!2^n}(\pi \he)^{1/4}}\left (
A\cos\left (\frac{\varphi}{2}\right )
H_n\makebox{\small $\left (\frac{x+\delta}{\sqrt{\he}}\right )$}
\e^{-\ff{1}{2\he}(x+\delta)^2}
-B\sin\left (\frac{\varphi}{2}\right )
H_n\makebox{\small $\left (\frac{x-\delta}{\sqrt{\he}}\right )$}
\e^{-\ff{1}{2\he}(x-\delta)^2}
\right ),
\\ \label{phi_n}
\tilde\phi_n(x)&=\frac{e^{\fitwo\theta(x)}}{\sqrt{n!2^n}(\pi \he)^{1/4}}\left (
A\sin\left (\frac{\varphi}{2}\right )
H_n\makebox{\small $\left (\frac{x+\delta}{\sqrt{\he}}\right )$}
\e^{-\ff{1}{2\he}(x+\delta)^2}
+B\cos\left (\frac{\varphi}{2}\right )
H_n\makebox{\small $\left (\frac{x-\delta}{\sqrt{\he}}\right )$}
\e^{-\ff{1}{2\he}(x-\delta)^2}
\right )
\end{align}
We should notice that if we take into account the general function $w(x)$ from the solutions of \eref{eq_diff},
the matrix $Q_0(x)$ would be instead 
\begin{align}
Q_0(x)&=
\left (
\begin{array}{cc}
x+iw(x)+\delta & 0\\
0 & x+iw(x)-\delta
\end{array}
\right )
\end{align}
and it is easy to show that the solutions $(\phi_n,\tilde\phi_n)$ transform into 
$(\phi_n,\tilde\phi_n)e^{i\he^{-1}\int^x w(y)dy}$.
We also define the currents $J_n(x)=\ftwo \he\Im (\phi_n^*(x)\partial_x\phi_n(x))$ and 
$\tilde J_n(x)=\ftwo \he\Im (\tilde\phi_n^*(x)\partial_x\tilde\phi_n(x))$. The sum of the 
currents satisfies
\begin{align}\nn
J_n(x)+\tilde J_n(x)&=\ftwo\omega(x)\left ( |\tilde\phi_n(x)|^2-|\phi_n(x)|^2 \right )
\end{align}
The function $\omega(x)$ is physically equivalent to an internal current, proportional
to the imbalance between the two orbitals occupancy.  
$A$ and $B$ are chosen such that for each orbital there are $\xn_0$ and $\tilde \xn_0$ electrons in the ground state: $\int\d 
x|\phi_0|^2=\xn_0$ and $\int\d x|\tilde\phi_0|^2=\tilde \xn_0$. We could also eventually choose $\xn_0$ and $\tilde \xn_0$ as function of $A$ and $B$. $A$ and $B$ satisfy the system of equations
\begin{align}\nn
|A|^2\cos^2\left(\frac{\varphi}{2}\right )+|B|^2\sin^2\left(\frac{\varphi}{2}\right )-
\cos\left(\frac{\varphi}{2}\right )\sin\left(\frac{\varphi}{2}\right )
(A^*B+AB^*)\e^{-\frac{\delta^2}{\he}}&=\xn_0,
\\
|A|^2\sin^2 \left(\frac{\varphi}{2}\right )
+|B|^2\cos^2 \left(\frac{\varphi}{2}\right )
+\sin\left(\frac{\varphi}{2}\right )\cos\left(\frac{\varphi}{2}\right )(A^*B+AB^*)\e^{-\frac{\delta^2}{\he}}&=
\tilde \xn_0
\end{align}
We can parametrize coefficients $A$ and $B$ such that
\begin{align}\label{AB}
&A=\sqrt{\xn_0+\tilde \xn_0}\cos(\phi)\e^{i\theta_A},
\;
B=\sqrt{\xn_0+\tilde \xn_0}\sin(\phi)\e^{i\theta_B}
\end{align}
where $\phi$ satisfies the following equation
\bb
\cos(2\phi)-\frac{g}{x_c}\cos(\theta_A-\theta_B)\sin(2\phi)\e^{-\delta^2/\he}=\frac{\xn_0-\tilde \xn_0}{\xn_0+\tilde \xn_0}
\frac{\delta}{x_c}
\ee
We can choose $\theta_A$ and $\theta_B$ such that $\cos(\theta_A-\theta_B)=0$ to remove the dependence of $\phi$
with $\he$. The resulting equation for $\cos(2\phi)$ has solutions provided that $|\xn_0-\tilde \xn_0|/(\xn_0+\tilde \xn_0)<x_c/\delta$.
This excludes configurations such that $\xn_0=1$ and $\tilde \xn_0=0$ for example, since $x_c/\delta<1$ for $g>0$, in
which case the two components of the wavefunction are "intricated": each quasiparticle in one
band has a non-zero component in the other band when $g>0$, due to the tunnel effect induced by the unitary rotation in \eref{eq_unitary} when $\varphi\neq 0$, even in absence of magnetic field. The critical ratio between the two densities when $\cos(2\phi)=\pm 1$ is $\xn_0/\tilde \xn_0=(\delta\pm x_c)/(\delta\mp x_c)$. This means that for a density $\xn_0$ in the first orbital, there is a transfer of a small component proportional to $g^2\xn_0$ to the second orbital when $g\ll 1$. 
For example, when $B=0$, the ratio between the two densities is simply
\bb
\frac{\tilde \xn_0}{\xn_0}=\tan^2\left (\frac{\varphi}{2}\right )=\frac{g^2}{x_c+\delta}
\ee
Unless $g=0$, in which case $\xn_0=|A|^2$ and $\tilde \xn_0=0$, the transmitted part $\tilde \xn_0$ is
proportional to $g^2\xn_0$.
When $\xn_0=\tilde \xn_0$, we will consider in the following the real solutions of the system (\ref{AB}) with $\theta_A=\theta_B=0$ and for which $\phi$ is explicitly given by
\bb
\tan(2\phi)=\frac{x_c}{g}\e^{\frac{\delta^2}{\he}}
\ee
If $\omega(x)$ is constant, the total integrated current vanishes in this case. 
In the limit of small coupling $g$, we obtain $|A|^2\simeq \xn_0$,
$|B|^2\simeq \tilde \xn_0$, and $\phi\simeq \pi/4$: the two functions $\phi_0$ and 
$\tilde\phi_0$ decouple as each electron is localized in his own orbital.
\subsection{Overlap function}
%
As a first application, we consider the case in \efig{fig1} for which $\omega(x)=1$ and define the overlap or transmission coefficient between the two wavefunctions (\ref{phi_n}) to evidence and study the oscillatory nature of the system as function of the inverse magnetic field. 
For this, we define the inner product or transfer function $T_{mn}=\int_{-\infty}^{\infty}dx\phi_m(x)\overline{\tilde\phi_n}(x)dx$ between the Landau states $m$ and $n$. After computing the product using integral formula for Hermite polynomials \cite{Prudnikov_vol2}, we obtain
\begin{align}\nn
T_{m\le n}&=\sqrt{\frac{m!2^n}{n!2^m}}(i\sqrt{\he})^{m-n}\frac{ge^{-1/\he}}{\delta}
\Big [
\left (\cos(\phi)^2e^{2i\delta/\he}-\sin(\phi)^2e^{-2i\delta/\he}
\right )L_m^{n-m}
 \left (\makebox{\footnotesize $\frac{2}{\he}$}\right )
\\ \nn
&-
e^{-\delta^2/\he}\sin(\phi)\cos(\phi)
\left ( \frac{g}{x_x+\delta}(1+i\delta)^{n-m}-\frac{x_x+\delta}{g}(1-i\delta)^{n-m}
\right )
L_m^{n-m}\left (\makebox{\footnotesize $\frac{2}{\he}(1+\delta^2)$}\right ) 
\Big ],
\\ \nn
T_{m\ge n}&=\sqrt{\frac{n!2^m}{m!2^n}}(i\sqrt{\he})^{n-m}\frac{ge^{-1/\he}}{\delta}
\Big [
\left (\cos(\phi)^2e^{2i\delta/\he}-\sin(\phi)^2e^{-2i\delta/\he}
\right )L_n^{m-n}
\left (\makebox{\footnotesize $\frac{2}{\he}$}\right ) 
\\ \nn
&-
e^{-\delta^2/\he}\sin(\phi)\cos(\phi)
\left ( \frac{g}{x_x+\delta}(1-i\delta)^{m-n}-\frac{x_x+\delta}{g}(1+i\delta)^{m-n}
\right )
L_n^{m-n}\left (\makebox{\footnotesize $\frac{2}{\he}(1+\delta^2)$}\right )
\Big ],
\\ \nn
T_{nn}&=\frac{ge^{-1/\he}}{\delta}
\Big [
\left (\cos(\phi)^2e^{2i\delta/\he}-\sin(\phi)^2e^{-2i\delta/\he}
\right )L_n
\left (\makebox{\footnotesize $\frac{2}{\he}$}\right ) 
\\ 
&+
\frac{x_ce^{-\delta^2/\he}}{g}\sin(2\phi)
L_n\left (\makebox{\footnotesize $\frac{2}{\he}(1+\delta^2)$}\right ) 
\Big ]
\end{align}
%
\begin{figure}%
\centering
\includegraphics[width=0.9\columnwidth,clip,angle=0]{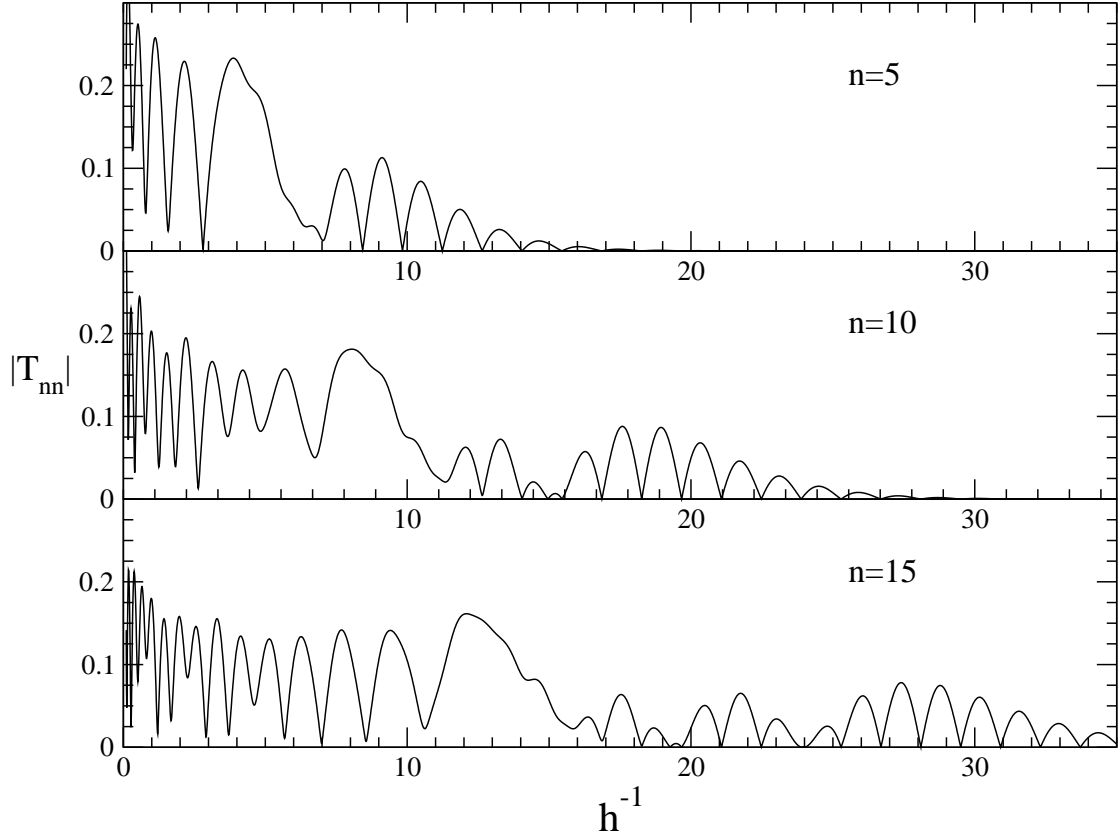}%
\caption{Overlap function modulus $|T_{nn}|$ as function of the inverse field $\he^{-1}$ for parameters $g=0.5$, $x_c=1$, and for $n=5,10,15$.}%
\label{fig3}%
\end{figure}
%
\begin{figure}%
\centering
\includegraphics[width=0.9\columnwidth,clip,angle=0]{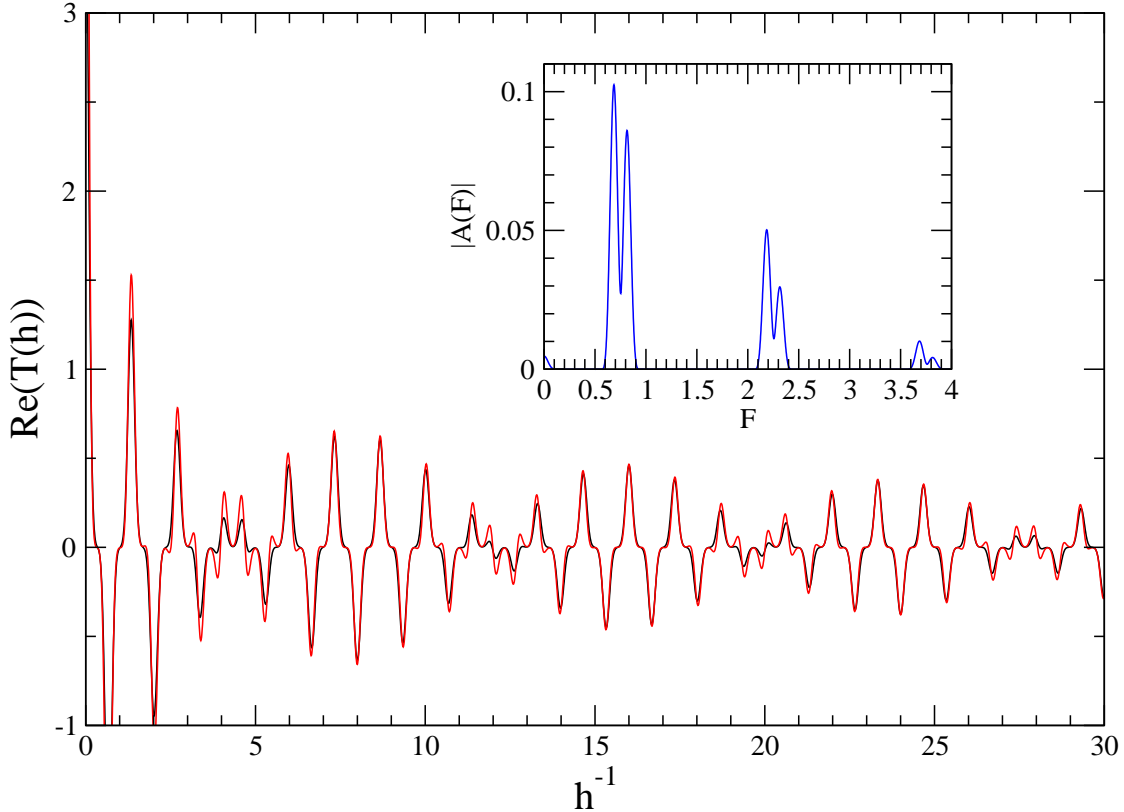}%
\caption{Real part of the overlap $T(\he)$ as function of $1/h$ with parameters $E=3$, $g=0.2$, $\omega(x)=1$, and $x_c=1$. Broadening parameter $\Gamma=0.1$. Red curve is the semiclassical expression (\ref{Thsc}).
Inset: Fourier transform (FT) of $\Re(T(\he)$. The first peak is estimated 
to be located at frequency $F=0.68$ and the second at $F=0.82$.}%
\label{fig3b}%
\end{figure}
%
%
where $L_m^{n-m}$ are generalized Laguerre polynomials. We define the transmission function $T(\he)$ at a given energy $E$
\bb\label{Th}
T(\he)=\sum_{n\ge 0}\delta\left (n-\frac{E}{2\he}\right )T_{nn}(\he)
\ee
In the numerical situation of \efig{fig3b}, we will use a broadening of the Landau levels with a Gaussian function of width $\Gamma\ll 1$ instead of the delta function $\delta(x)\rightarrow \e^{-x^2/2\Gamma^2}(2\pi\Gamma^2)^{-1/2}$.
Oscillations as function of the inverse field $B^{-1}$ are characterized by their frequencies $F$ in the Fourier spectrum. For example, in the de Haas-van Alphen semi-classical theory, the Fourier spectrum of the magnetization oscillations are given by a set of discrete fundamental frequencies $F$ and their harmonics that are proportional to the surface area $\Ai$ of all the closed cyclotronic orbits in the Fermi surface $(x,p)$ at a given energy: they are equal to $F=\Ai \hbar/2\pi e$. In term of inverse effective field variable $\he^{-1}$, they are expressed instead as $F=\Ai/2\pi$. In absence of coupling $g=0$ and with $\omega$ constant, we can estimate the geometrical expressions of the surface areas $\Ai_{\alpha}$ enclosed by the blue curve and $\Ai_{\beta}>\Ai_{\alpha}$ in red on \efig{fig1} which are given respectively by
\begin{align}\nn
\Ai_{\alpha}&=2E\arcsin\left (\sqrt{\frac{E-\omega^2-x_c^2}{E}}\right )
-2\sqrt{(E-\omega^2-x_c^2)(\omega^2+x_c^2)},
\\
\Ai_{\beta}&=2\pi E-\Ai_{\alpha}=2\Ai_0-\Ai_{\alpha}
\end{align}
where $\Ai_0=\pi E$ is the area of the two circular cyclotronic orbits in the non interacting case, with radius $\sqrt{E}$. In \efig{fig3b} is plotted the real part of the overlap $T(\he)$ as function of $1/h$ with parameters $E=3$, $x_c=1$, and $g=0.05$. We observe that the amplitude of $T(\he)$ decreases with $\he$ in the semiclassical limit. The presence of two frequencies, one small and one large, is observed with beating effect in the field regime considered. With these parameters and $g=0$, we have the de Haas-van Alphen frequency $F_{0}=E/2=1.5$ \footnote{The de Haas-van Alphen theory for magnetization applied to the linear spectrum $E_n=2hn$ gives the dominant frequency $F_0=\Ai_0/2\pi$.}, in addition to $F_{\alpha}\simeq 0.138$, $F_{\beta}\simeq 2.862$. We also assume that the areas are still approximately given by these values when $g$ is small. We then compute the Fourier amplitude of $T(\he)$ with respect to $\he^{-1}$
\begin{align}\label{FT}
A(F)=\int_0^{\infty} dh^{-1}T(\he)\e^{-2i\pi F/h}=\frac{1}{F_0}\sum_{n\ge 1}
T_{nn}(F_0/n)e^{-2i\pi n F/F_0}
\end{align}
In inset of \efig{fig3b}, we have plotted $|A(F)|$ which presents double peaks at regular locations
we identify to be equal $nF_0+(F_0\pm F_{\alpha})/2$, with $(F_0-F_{\alpha})/2\simeq 0.68$
and $(F_0+F_{\alpha})/2\simeq 0.82$. Each pair of peaks is separated by $F_0$, and the beating frequency is 
$F_{\alpha}$. $F_{\alpha}$ can be considered physically as the overlap area between the two orbits and therefore
characterizes the function $T_{nn}(\he)$ at small coupling $g$, whereas $F_0$ represents the periodicity of the energy density. 

In the semiclassical regime $\he\ll 1$, $T(\he)$ can be found analytically with the saddle point method. Using the representation integral of the delta function, one may consider the following function relevant to compute $T(\he)$:
\begin{align}\label{GE}
\GE(\he,\zd)=\e^{-\zd/\he}\int_{-\infty}^{\infty}\frac{dx}{2\pi}\sum_{n\ge 0}
L_n\left (\makebox{\footnotesize $\frac{2\zd}{\he}$}\right )\e^{ix(n-E/2\he)}
\end{align}
with $\zd>0$. Using the generating function for the Laguerre polynomials, we can perform the series and obtain
formally~\footnote{We need normally a small parameter $0<\epsilon\ll 1$ such that $\e^{i(x+i\epsilon)}<1$
in order for the series in \eref{GE} to be convergent, and set $\epsilon=0$ after the calculations.}
\begin{align}
\GE(\he,\zd)=\e^{-\zd/\he}\int_{-\infty}^{\infty}\frac{dx}{2\pi}\frac{\e^{-i\varphi(x)/\he}}{1-\e^{ix}},
\;\varphi(x)=\ftwo Ex-\frac{2i\zd}{\e^{-ix}-1}
\end{align}
We then apply the saddle point method to search for the 
solutions of $\varphi'(x)=0$. We find that the solutions are given by the equation
$\sin(x/2)^2=\zd/E$ or $x=\pm 2\arcsin(\sqrt{\zd/E})+2\pi k$, with $k$ integer.
Using the second derivative $\varphi''(x)=\ftwo\zd\cos(x/2)/\sin(x/2)^3$, we obtain, after some algebra,
the following behavior
\begin{align}\nn
\GE(\he,\zd)&\simeq \sqrt{\frac{\he}{\pi}}\frac{1}{(\zd(E-\zd))^{1/4}}
\sin\left (2\pi \frac{\Fc{\zd}}{\he}+\frac{\pi}{4}+\phase{\zd}\right )\drho(\he),
\\
\drho(\he)&=\sum_n\delta\left (n-\frac{E}{2\he}\right ),\;
\tan\phase{\zd}=\sqrt{\frac{\zd}{E-\zd}}
\end{align}
The frequency $\Fc{\zd}$ is given by
\bb
2\pi\Fc{\zd}=E\arcsin\left (\sqrt{\frac{\zd}{E}}\right )+\sqrt{\zd(E-\zd)}
\ee
From this expression, $T(\he)$ is given in this limit by

\begin{align}\label{Thsc}
T(\he)&\simeq
\Big [\frac{g}{\delta}
\left (\cos(\phi)^2e^{2i\delta/\he}-\sin(\phi)^2e^{-2i\delta/\he}
\right )\GE(\he,1)+
\frac{x_c}{\delta}\sin(2\phi)
\GE(\he,1+\delta^2)
\Big ]\drho(\he)
\end{align}
The Fourier spectrum should therefore present a series of peaks at harmonics of $F=F_0$ due to the 
periodicity of $\drho(\he)$, and two beating frequencies $\Fc{1}$ and $\Fc{1+\delta^2}\simeq(F_0-F_{\alpha})/2$, the latter being dominant in amplitude when $g$ is small. The red curve in \efig{fig3b} represents the expression
(\ref{Thsc}) which is in good agreement with the solutions given by \eref{Th} for the field range.
%
\subsection{Charge transfer oscillations}
%
The second example of application concerns the computation of the electron density or charge density for each orbital $\xn_n=\int\d x|\phi_n|^2$ and $\tilde \xn_n=\int\d x
|\tilde \phi_n|^2$, as function of the Landau level $n$. We can, as before, integrate over Hermite polynomials and obtain
\begin{align}\nn
\xn_n&=\xn_0+g\frac{(\xn_0+\tilde \xn_0)}{2\delta}\sin(2\phi)
\left (1-L_n\left (\makebox{\footnotesize $\frac{2\delta^2}{\he}$}\right )\right )\e^{-\delta^2/h}
\\
\tilde \xn_n&=\tilde \xn_0-g\frac{(\xn_0+\tilde \xn_0)}{2\delta}\sin(2\phi)
\left (1-L_n\left (\makebox{\footnotesize $\frac{2\delta^2}{\he}$}\right )\right )\e^{-\delta^2/h}
\end{align}
The sum of the densities is conserved but there is a transfer of charge as function of index $n$.
%
%
\begin{figure}%
\centering
\includegraphics[width=0.9\columnwidth,clip,angle=0]{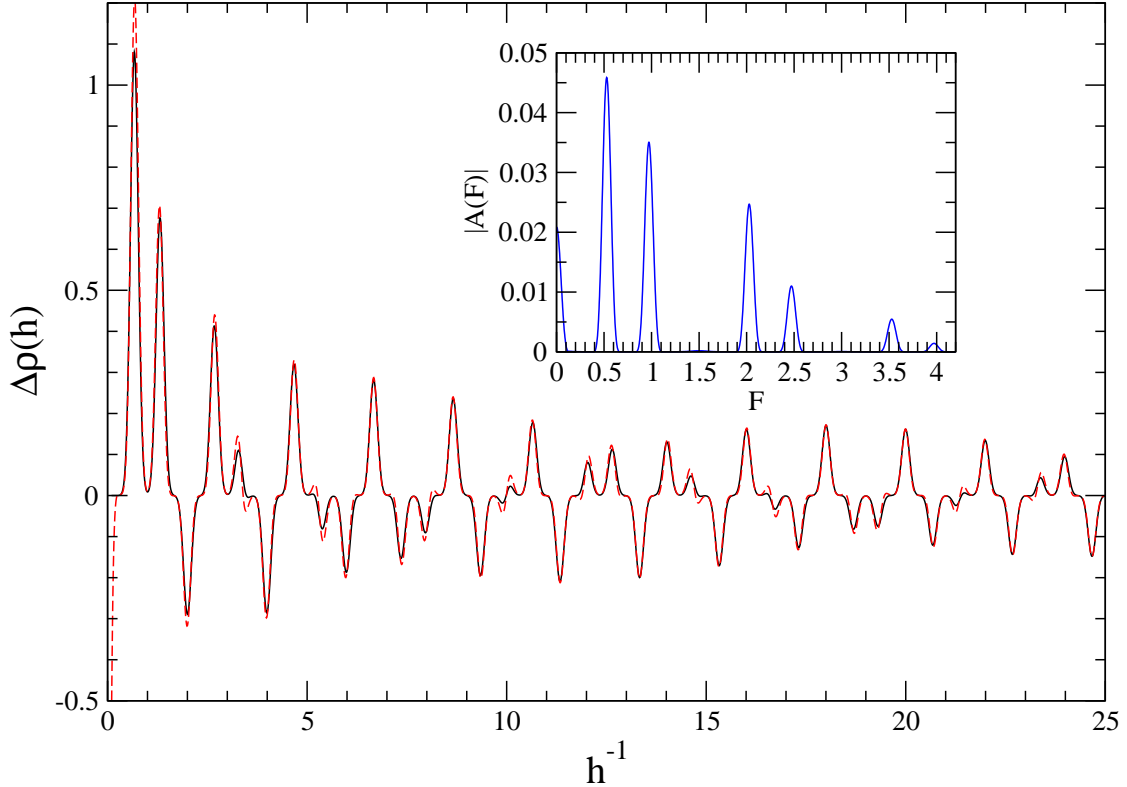}%
\caption{Density fluctuations $\DN(\he)$, \eref{DN}, as function of inverse fields $\he^{-1}$
(black line) for the parameters $\xn_0=\tilde \xn_0=1$, $E=3$, $g=0.2$, and $x_c=1$.
Broadening parameter $\Gamma=0.1$. Red line is the semiclassical result (\ref{DNcl}). Inset: Fourier amplitude
of the black curve. Fundamental frequency is located at $F=0.527$, which corresponds to $\Fc{\delta^2}$ (see text),
and harmonics at $nF_0\pm \Fc{\delta^2}$ with $n$ integer.}%
\label{fig4}%
\end{figure}
%
We can consider the transfer charge function (we set $\xn_0=\tilde \xn_0=1$ in the following)
\begin{align}\label{DN}
\DN(\he)=\sum_{n\ge 0} \delta\left (n-\frac{E}{2\he}\right )(\xn_n-\tilde \xn_n)
\end{align}
and study as before the semi-classical limit $\he\ll 1$, using function $\GE(\he,\delta^2)$, \eref{GE}.
We obtain the expression for $\DN(\he)$ in this limit
\begin{align}\label{DNcl}
\DN(\he)\simeq\frac{2g}{\delta}\sin(2\phi)
\left [\e^{-\delta^2/h}-
\sqrt{\frac{\he}{\pi}}\frac{1}{\sqrt{\delta}(E-\delta^2)^{1/4}}
\sin\left (2\pi \frac{\Fc{\delta^2}}{\he}+\frac{\pi}{4}+\phase{\delta^2}\right )
\right ]\drho(\he)
\end{align}
$\DN(\he)$ presents clearly beating oscillations with frequency $\Fc{\delta^2}$.
For $\xn_0=\tilde \xn_0=1$, we have plotted in \efig{fig4} the density fluctuations as function of the inverse
field, for parameters $E=3$, $g=0.2$, and $x_c=1$. One finds the beating frequency $\Fc{\delta^2}\simeq 0.527$ which is 
in agreement with the Fourier spectrum plotted in inset of \efig{fig4}.

\section{General case}
%
%
We generalize in this section the model of two oscillators by considering the following extension of the matrix $Q(x)$ in \eref{Qgen} representing a system of $2(\nm+1)$ interacting orbits, where $\nm\ge 0$,
\bb\label{QN}
Q(x)=
\left (
\begin{array}{cccc}
\makebox{\footnotesize $x+(2\nm+1)x_c+i\omega(x)$} & g\e^{-i\theta(x)} & 0 \cdots &\\
g\e^{i\theta(x)} & \makebox{\footnotesize $x+(2\nm-1)x_c-i\omega(x)$} & g\e^{i\theta(x)} & 0  \cdots \\
\cdots &  & &   \\
\cdots & 0 & g \e^{i\theta(x)} & \makebox{\footnotesize $x-(2\nm+1)x_c-i\omega(x)$}
\end{array}
\right )
\ee
%
\begin{figure}%
\centering
\includegraphics[width=0.9\columnwidth,clip,angle=0]{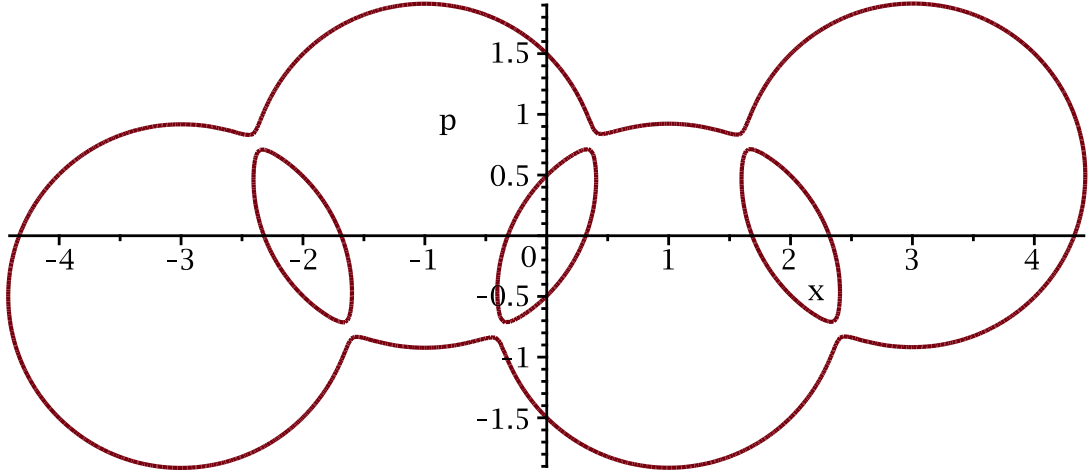}%
\caption{Fermi surface of \eref{QN} with $\nm=1$, representing 4 coupled orbits with parameters $E=2$, $g=0.1$, $\omega(x)=1/2$, and $x_c=1$.}%
\label{fig5}%
\end{figure}
%
This extended matrix satisfies the commutation relation $P(x)=1$ provided that $\theta(x)=2\he^{-1}\int^x\omega(y)dy$ as before. There are alternating internal current generated by $\pm \omega(x)$ in each orbital. For $\nm=1$, we have plotted in \efig{fig5} the classical Fermi surface composed of a linear chain containing 4 orbits. We can compute the ground state by solving the differential equation $\D\vk_0=0$. Noting the $(2\nm+2)$-component vector $\vk_0=(\phi_0^{(\nm)},\cdots,\phi_0^{(-\nm-1)})^\trans$, the system of equations $\he\partial_x\vk_0+Q(x)\vk_0=0$ can be reduced by the transformation
$\vk_0\rightarrow \tilde\vk_0$ where 
\bb\nn
\tilde \vk_0=(\tilde\phi_0^{(\nm)},\cdots,\tilde\phi_0^{(-\nm-1)})^\trans,\;
\vk_0=\e^{-\frac{x^2}{2h}}\diag(\e^{-\fitwo\theta(x)},\e^{\fitwo\theta(x)},\cdots,\e^{\fitwo\theta(x)})\tilde \vk_0
\ee 
This simplifies the system of differential equations $-\he\partial_x\tilde\vk_0
=\tilde Q \tilde \vk_0$, where $\tilde Q$ is the constant matrix:
\bb\label{QN0}
\tilde Q=
\left (
\begin{array}{cccc}
\makebox{\footnotesize $(2\nm+1)x_c$} & g & 0 \cdots &\\
g & \makebox{\footnotesize $(2\nm-1)x_c$} & g & 0  \cdots \\
\cdots &  & &   \\
\cdots & 0 & g & \makebox{\footnotesize $-(2\nm+1)x_c$}
\end{array}
\right )
\ee
We can diagonalize this matrix by searching for the eigenvectors $v_k^{\lambda}$, with $k=-\nm-1,\cdots,\nm$
\bb
v_{k-1}^{\lambda}+v_{k+1}^{\lambda}+\frac{(2k+1-\lambda)x_c}{g}v_k^{\lambda}=0
\ee
whose solutions are given by Bessel functions
\bb
v_k^{\lambda}=aJ_{\mu-k}(g/x_c)+bY_{\mu-k}(g/x_c)
\ee
where we have defined $\lambda=2\mu+1$.
We impose as boundary conditions $v_{\nm+1}^{\lambda}=v_{-\nm-2}^{\lambda}=0$. This leads to the 
equation for the discrete eigenvalues $\mu$
\bb\label{wronskian}
J_{\mu+\nm+2}(g/x_c)Y_{\mu-\nm-1}(g/x_c)=Y_{\mu+\nm+2}(g/x_c)J_{\mu-\nm-1}(g/x_c)
\ee
There are $2(\nm+1)$ solutions $\lambda_k=2\mu_k+1$ to this equation which is a polynomial in $\mu$: $\mu_{\nm}>\cdots>\mu_{-\nm-1}$ (for small $g$, we have the approximation $\mu_k\simeq k$). $x=-\lambda_k$ represents the position of the $k^{th}$ oscillator on the chain. Then we can choose the following eigenvectors:
\bb
v_k^{\lambda}=\frac{Y_{\mu-\nm-1}(g/x_c)J_{\mu-k}(g/x_c)-J_{\mu-\nm-1}(g/x_c)
Y_{\mu-k}(g/x_c)}{Y_{\mu-\nm-1}(g/x_c)J_{\mu-\nm}(g/x_c)-J_{\mu-\nm-1}(g/x_c)
Y_{\mu-\nm}(g/x_c)}
\ee
For example, for $\nm=1$, \eref{wronskian} leads to the quartic equation for $\mu$
\bb
16\mu^4+32\mu^3-4\{ 3(g/x_c)^2+4\}\mu^2-4\{3(g/x_c)^2+8\}\mu+(g/x_c)^4=0
\ee
It is then easy to show that the components of $\vk_0$ can be expressed as
\bb
\phi_0^{(k)}(x)=\e^{\pm \fitwo\theta(x)}\sum_{l=-\nm-1}^{\nm}C_lv_k^{\lambda_l}\e^{-\frac{(x+\lambda_lx_c)^2}{2\he}}
\ee
where $C_k$ are constants determined by the $(2\nm+2)$ conditions $\int\d x|\phi_0^{(k)}|^2=\xn_0^{(k)}$ for the 
density conservation. The sign of the wavefunction phase $\theta(x)$ is alternating depending on the oscillator position, with a minus sign for the first oscillator located at $x=-\lambda_{-\nm-1}$.
The excited states $\phi_{n\ge 0}^{(k)}(x)$ with energy $E_n=2hn$ are then constructed by identification to \eref{phi_n} with Hermite polynomials
\bb
\phi_n^{(k)}(x)=\frac{\e^{\pm \fitwo\theta(x)}}{\sqrt{n!2^n}}
\sum_{l=-\nm-1}^{\nm}C_lv_k^{\lambda_l}H_n\makebox{\small $\left (\frac{x+\lambda_lx_c}{\sqrt{\he}}\right )$}\e^{-\frac{(x+\lambda_lx_c)^2}{2\he}}
\ee
%
\section{Conclusion}
%
In this paper we have shown that we can construct exact solutions of interacting bands in a magnetic field, from
a factorized Hamiltonian and using a unitary transformation acting on independent band orbitals,
see \eref{eq_unitary}. The simple extension of the model leads to the matrix structure, \eref{QN}, which is a Fermi surface
composed of several coupled orbits and allows us to determine the eigenvectors. In particular for the simple case of two coupled orbits, we have evaluated the overlap function, \eref{Th}, whose oscillating behavior contains the geometrical characteristics of the cyclotronic trajectories on the Fermi surface. We also considered the charge transfer between the two orbitals, and showed that it oscillates as function of the inverse field with a beating frequency $\Fc{\delta^2}$, in addition to the dominant de Haas-van Alpen frequency of the Landau energy spectrum.


\begin{acknowledgments}

This work was supported by the Brain Pool Program through the National Research Foundation of Korea (NRF-2018H1D3A2065321). I want to thank the referee for useful comments.

\end{acknowledgments}
\bibliography{tunneling}

\end{document}